# Infected surfaces of vehicles as possible way of people's infection by bird flu pathogenic culture

*Possible variant of people's infection by bird flu pathogenic culture in passing of everyday infection is presented in the work: through the contact of open parts of the skin with infected surfaces of the vehicle, that is the sequent of the reused water, which contains all species spectrum of pathogen accumulated on the urban areas, used in process of washing.*

**Manuylov M.B.** – Candidate of technical sciences, Research director of "EKOS" company, Kharkov, Ukraine
**nc-ekos@ngs.ru**

**Mavrov I.I.** – Doctor of medical sciences, Director Research Institute of Venereology and Dermatology, Kharkov, Ukraine

**Moskovkin V.M.** - Doctor of Geographical Science, Deputy of Vice-Rector on innovation activity, Belgorod State University, Russia
moskovkin@bsu.edu.ru

The real threat of beginnings of planetary scope catastrophe exists – it's the possibility of pandemic repetition of 1918 called "Spaniard", which had taken the lives of 50 million men. In many scientific' opinions, also the experts of World Health Organization (WHO) hold this version, "Spaniard" virus is the mutative form of bird flu pathogenic virus, which had the possibility to be communicated from person to person by respiratory way. It is natural, that at present time the quantity of victims will be increased with the advent of present viral culture. There are real prerequisites for it: in comparison with the beginning of the last century the intensity of people's moving on the planet rose steeply, the speeds of such moving increased, the possibilities of visits of almost all countries of the world exist.

Main task, formulated by WHO, is the maximum decrease of probability of people's infection by bird flu pathogen viruses in all over the world, which minimize the possibilities of mutative forms beginnings the most dangerous for people. The solving of this problem means the search of people's infection possible ways and solving on its removal, information of all tellurians about necessary safeguards, monitoring and studying of all disease cases and something else.

Before coming to version argumentation, called in the name of the work, we will formulate the knowledge database based on public information giving by WHO through mass media.

1. **The way of birds' infection: existent information about cause-effect relations.**

    Variant № 1-1. Dung of ill bird or bird-carrier of virus, gets to reservoir, at the same time finely divided suspended matter comprising viruses are generated. Infection of health birds got into this reservoir happens through alimentary canal.
    Conclusion:
    - infection of the reservoirs by bird flu pathogen viruses comprising in suspended matter is possible;
    - Bird flu virus can keep vitality in aquatic environment for some period of time.

    Variant № 1-2. Dung of ill bird or bird-carrier of virus gets into the feed or on ground surface. Infection of health birds through alimentary canal takes place.
    Conclusion:
    - Bird flu virus keeps vitality; be in ground surface for a long time.

2. **People's infection way: information about cause-effect relations exists.**

    Variant № 1-2. Infection through alimentary canal: use of bird infected meat or getting of the virus from infected reservoir into human organism.
    Conclusion: probability of people's infection is minimal:
    - heat treatment of bird carcass kills the virus and people are informed about safeguards;
    - time of birds migration, the period of maximum danger, doesn't coincide with the period of reservoir recreation use;
    - the technology of drinking water production from surface water provides for many-stage refinement and disinfection – the possibility of people's infection tends to zero.

    Variant № 2-2. Aerogenic way of infection: by the works with dry birds dung precipitated particulate pollutant infected by bird flu virus can be generated.
    Conclusion: probability of people's infection is minimal:
    - mucous membrane of nose and nasopharynx delay almost all particles of aerosols, diameter of which is equal or more 10 mkm; and approximately 50% of particles, size of which is from 1 till 5 mkm; in tracheobronchitis tree, tube of which is represented by trachea, bifurcation of bronchus leads to the respiratory bronchioles, which divides into alveolar ductules, opens in atria and circled by alveoles; bronchial area is abundantly supplied with lymphatic and lymphoid tissue, viruses can direct penetrate into lymphatic;
    - chemical composition and bird dung structure don't allow to form the particles with the size less than 20 mkm, particles formation of 5 mkm and less has the randomicity and is due to the fact of mechanical destroy of single aggregates.

Variant № 2-3. Virus bringing in circulatory system: cutting of infected bird carcass, under which is possible the infection of people through the gate of infection: grazes, cuts, microtraumas and other.

Conclusion: probability of people's infection is small:
- sell of infected bird is small;
- the population of all countries is informed about necessary safeguards by the work concerned with the birds cutting and cooking.

3. **Consider the knowledge database characterized the fact of everyday people infection: infection is happened in case of touch of open parts of skin with infected pathogenic germ of the vehicles surfaces.**

**3.1 The reasons of reused waters microbiological infections and surfaces of vehicles.**

The formation of road sweeping characterized by difficult microbiological composition with obligatory presence of pathogenic bacterium and viruses wide spectrum happens on the urban areas [1,2]. During the moving of transport, accumulation of charged road sweeping on the bottoms and transfer of pollution in vehicles on the shoes happens (on this stage the probability of people infection is tended to zero) [3,4]. In washing process of vehicles, wash-out of infected sweeping and conservation of microorganisms in system of treatment plant preparation of reused water is realized [5].

At the present time the quality of reused waters on the territory of Ukraine is regulated by Standard Norms and Rules 11-9374: concentration of suspended matters – under 40 mg/l; mineral oils – under 15 mg/l; tetraethyl lead – 0,001 mg/l, guideline about microbiological composition are absent. Developing the technologies of water recycling, the specialists follow Standard Norms and Rules guideline, that's why typical treatment plants consist of settling and coarse filter. We can give an example: granulometric composition of suspended matters contains in effluent (reused water), which was formed in the vehicle washing process is given in table 1. Wastewater passed the refinement in flowing settling and on the filter "Avtopen" (development of VNIIVO and VODGEO Institutes (Kharkov, Ukraine)).

It is significant that the quality of received reused water exceeds the demands of Standard Norms and Rules: concentration of suspended matters 27,5 mg/l, mineral oils 9 mg/l, tetraethyl lead – less than 0,001 mg/l [6], and percentage distribution of particles by sizes similarly to spectrums of sedimentary aerosols of industrial and vehicle origin [7, 8].

**Table № 1.**
**Averaged granulometric composition of suspended matters contain in recycling water of vehicles washing.**

| Range of particles sizes, mkm | Midium particles size, mkm | Particles distribution on fractions, % |
|---|---|---|
| 0, 2 – 5,0 | 2,6 | 26,05 |
| 5, 0 – 10,0 | 7,5 | 14,46 |
| 10,0-15,0 | 12,5 | 12,25 |
| 15,0-20,0 | 17,5 | 10,28 |
| 20,0-25,0 | 22,5 | 9,52 |
| 25,0-30,0 | 27,5 | 6,44 |
| 30,0-35,0 | 32,5 | 5,48 |
| 35,0-40,0 | 37,5 | 4,26 |
| 40,0-45,0 | 42,5 | 3,17 |
| 45,0-50,0 | 47,5 | 2,78 |
| 50,0-55,0 | 52,5 | 2,14 |
| 55,0-60,0 | 57,5 | 1,42 |
| 60,0-65,0 | 62,5 | 1,01 |
| 65,0-70,0 | 67,5 | 0,74 |
|  |  | 100,00% |

As we can see from the data given in table № 1 diapason of particles is limited to upper level 67,5 mkm, using other technologies of refinement, which follow the demands of Standard Norms and Rules – concentration of suspended matters 40 mg/l, maximum sizes of particles form from 85 till 100 mkm. It's natural as disinfection doesn't take place; all species composition of microorganisms, initially present in wastewater and road sweeping of urban areas, keeps in reused water.

The use of infected waters in washing process comes to vehicle surfaces infection, during touching with it, people infection happens – the probability of infection depends on pathogenic bacteria and viruses virulence.

**3.2. The population exposed to influence of concerned factor of everyday infection.**

All the people live in Ukraine, people transit the border, tourists come for rest exposed risk factor from everyday infection, because vehicle washing on the territory of Ukraine: transit vehicle, railway and trams; buses and trolley buses; cars and vehicles, is realized only with the use of water recycling.

Present approach to vehicle washing is typical for all CIS countries, as common legal rules of USSR – Standard Norms and Rules 11-93-74 and others.

**4. Algorithm of possible people's infection by bird flu pathogenic culture of concerned way of everyday infection.**
**4.1. Virus infection of urban area road sweeping.**

Migration routes of migration birds pass and over urban areas, exactly in cities of Germany, Czechia, Norway, Israel and many other countries 2006 were found the bird - carrier of virus pathogenic culture.

Conclusion:
- infection of urban areas road sweeping by dung of the birds-carrier of virus pathogenic culture in the form of local centre (area) of infection spread is possible;
- the probability of its event is high – big quantity of migrant bird's migration streams passes over the territory of Ukraine;
- for a long time virus keeps vitality in bird's dung, which can be the part of road sweeping – variant № 1-2.

### 4.2. Water infection of recycling waters systems of vehicle washing.

The existence of urban areas infection local centers, with big probability, will come to the transfer of pathogenic virus in treatment plants of water recycling system with obligatory infection of washing water.

It's caused:
- Clamminess of not dry bird's dung is very big – viscous on N.A. Kachinskiy's classification that increases the probability of pick up on the vehicle bottoms; pickup on the shoes with further transfer to vehicles;
- The probability of vehicle contact as well as the citizens with local centers of infected road sweeping is very high, that stipulates for long period of virus vitality and high intensity of vehicles and people moving on urban areas;
- Infected road sweeping washout and bird flu viruses' conservation in treatment plant settling happen in washing process of vehicles – variant № 1-1.
- Existent treatment plants of water recycling system provide the removal of particles with the sizes more than 67,5 mkm (point 3.1), in aquatic environments bird's dung forms aggregates in the range from 20 till 40 mkm.

Conclusion:
- transfer and conservation of infected road sweeping in treatment plants of water rotation system is possible;
- infection of sewage and treated water by the bird's dung particles, which contain bird flu pathogenic culture is possible.

### 4.3. Infection of vehicle surfaces in washing process.

Vehicles' washing is made with infected reused water (point 3-1) with possible content of bird dung particles involving bird flu pathogenic viruses. In connection with limit viscosity of bird dung particles, the probability of its fixation on vehicle surfaces tends to one.

Conclusion:

- in case of bird flu pathogenic culture presence in reused water, the probability of vehicle surfaces infection tends to one.

**4.4. Possible people's infection by bird flu pathogenic culture on of concerned way of everyday infection.**

People's infection can happens in case of touch of amazed parts of skin with infected vehicle surfaces – variant № 2-3; bringing in of virus in blood through the gate of infection – cuts, grazes, chaps, micro traumas and other**.**
Conclusion:
- the probability of people's infection occurs at infection of vehicle surfaces by bird flu pathogenic cultures.
This probability is in depending on:

4.4.1. Quantity, total area and place of formation of road sweeping infection centre on urban areas;

4.4.2. Volumes of suspended matters, which were formed from bird dung particles with content of virus pathogenic culture, those define thickness of its distribution on washed vehicle surfaces identifying the probability of contact of open skin parts with virus;

4.4.3. Users' volumes of vehicle with infected surfaces, volumes of risk group: we can choose children – cuts on the hands are often and women – regular care of hands (manicure) – as a result of it, microtrauma forms, that can be the gate for virus infection.

As at the present time in Ukraine official data about urban and other areas infection with bird flu pathogenic culture are absent, then as illustrative example of people's infection on present algorithm we can turn to more studied theme – tuberculosis:
- In Ukraine is tuberculosis epidemic, that means great quantity of people-carrier of the present disease; phlegm, slime and other people-carrier come to surfaces of waterproof coating and serve as sources of road sweeping infection – the formation of local centers happens; for a long time tubercle bacillus keeps its vitality in phlegm and others, which are the constituents of road sweeping (point 4.1.);
- Hundreds and thousands of infection centers form everyday on urban areas, that, with probability tends to one, comes to transfer of tubercle bacillus in treatment plants of water recycling systems with infection of sewage as well as treated water [1,5] (point 4.2);
- Infection of all washed surfaces of vehicles happens in the washing process, it is confirmed by the results of full-scale investigations: tubercle bacillus was present in treated reused water and on washed surfaces [1] (point 4.3.)
- People's infection is possible through the gate of infection (4.4) that makes contribution in distribution of tuberculosis epidemic in Ukraine.

Even if the present algorithm of people's infection is improbable, from the point of view of distribution of bird flu pathogenic cultures, then sufficiently many infectious diseases which transmitted with the help of concerned factor of everyday infection stays.

However, we must note, that the possibility of people's infection on mentioned algorithm of infection is marginally confirmed by WHO's experts: it's recommended to use the blocking mask or respirator during the care of poultry (prevention from infection by variant № 2-2), and to do the cleaning works of poultry-yards only in gloves and in protection suit and shoes, i.e. there is the possibility of infection by touch of open parts of skin with bird dung containing bird flu pathogenic cultures (prevention from infection by variant № 2-3).

In conclusion we note: minimization of probability of people's infection by bird virus pathogenic cultures is necessary, it's not necessary to turn to apocalypses described by many writers, for example Jack London "Crimson plague" [10] to realize it. Solving of the problem mentioned in the article is very simple: vehicle washing must be made with water that not contain pathogenic germs – drinking or industrial water, at least – to make disinfection of surfaces washed with infected water.

**Warning:**

Mentioned algorithm of people's everyday infection has one more problem – clean surface of vehicle is psychologically taken as safe and we don't take necessary measures of protection that can lead to undesirable consequences.